\documentclass[twocolumn,prb]{revtex4}
\usepackage{epsfig}
\usepackage{amsmath}
\usepackage{amssymb}
\usepackage{graphicx}
\usepackage{multirow}
\usepackage[usenames]{color}
\usepackage[english]{babel}

\begin{document}
\title{Multiferroic BiFeO$_3$-BiMnO$_3$ Nanocheckerboard From First Principles}         
\author{L. P\'alov\'a}
\author{P. Chandra}
\author{K. M. Rabe}
\affiliation{Center for Materials Theory, Department of 
Physics and Astronomy, Rutgers University, 
Piscataway, NJ 08854}
\begin{abstract}
We present a first principles study of 
an unusual heterostructure, 
an atomic-scale checkerboard of BiFeO$_3$-BiMnO$_3$,
and compare its properties to the two bulk constituent materials,
BiFeO$_3$ and BiMnO$_3$.
The ``nanocheckerboard'' is found to have a multiferroic ground state 
with the desired properties of each constituent: 
polar and ferrimagnetic due to BiFeO$_3$ and BiMnO$_3$, respectively.
The effect of B-site cation ordering on magnetic ordering in
the BiFeO$_3$-BiMnO$_3$ system is studied.
The checkerboard geometry is seen to give rise to a
a novel magnetostructural effect 
that is neither present
in the bulk constituent materials, nor
in the layered BiFeO$_3$-BiMnO$_3$ superlattice.
\end{abstract}

\maketitle

\section{Introduction}

Artificially structured oxides present intriguing opportunities for
material design. 
With dramatic advances in epitaxial growth techniques allowing atomic-scale 
control, 
experimental and theoretical attention has focused on 
strained-layer 
superlattices~\cite{Ohtomo02,Neaton03,Ahn04,Fong04,Johnston05,Dawber05b,Karin05,Dawber05,Lee05,Nakhmanson06}.
Properties significantly different from those in the bulk
have been observed,
leading to the possibility of designing new materials
at the nanoscale with enhanced functionalities ~\cite{Ramesh02,Cen09,Mannhart10}.
Recently, progress has been reported in the synthesis
of artificially structured oxides with lateral 
``nanocheckerboard'' (or nanopillar) patterning.
In particular, 
the length scale of this checkerboard ordering 
can be controlled by synthetic processes
and stoichiometry, offering promise for applications 
such as ultrahigh-density magnetic recording
media~\cite{Zheng04,Driscoll08,Yeo06,Zhang07a,Zhang07b,Guiton07}.

One functionality of particular current interest 
is multiferroicity: 
the combination of ferromagnetism and 
ferroelectricity, with coupling between the spontaneous 
polarization and the magnetization.
Room temperature multiferroic materials
with high magnetoelectric couplings
are desirable,
because they can support 
novel functionalities in 
electronic devices~\cite{Scott07,Vaz10}.
Magnetostructural and magnetoelectric couplings 
have been observed in 
a number of materials, including 
bulk~\cite{Kozlenko03,Lee09} and layered~\cite{Murata07} manganites,
epitaxial EuTiO$_3$~\cite{Fennie06},
EuSe/PbSe$_{1-x}$Te$_x$ 
multilayers~\cite{Lechner05},
and SrRuO$_3$/SrTiO$_3$ oxide interfaces~\cite{Rondinelli08}.

Because of the distinct natures of ferroelectric and ferromagnetic
ordering,
it has proved difficult
to find a single-phase 
room temperature multiferroic material with large polarization, large magnetization, and large magnetoelectric and/or magnetoelectric 
coupling~\cite{Hill00}.
Most current multiferroic devices are based
on nanocomposites~\cite{Nan08,Vaz10},
and advances in the
synthesis of artificially structured materials
further support studies of novel multiferroic heterostructures~\cite{Ramesh07}.
Exploring
the coupling of ferroelectric
and/or magnetic states to strain
has shown to be exceptionally fruitful
in many multiferroic nanocomposites~\cite{Nan08}.
The challenge is to anticipate
what new properties can arise in such heterostructures
from combining two distinct materials,
and how these properties depend on the geometry of the combination.

First-principles approaches are 
ideally suited for meeting this challenge.
These methods allow searching 
over a variety of compositions, heterostructure geometries, and 
structure types to
find a material 
with the desired properties~\cite{Baettig05}.
With first-principles methods, it is possible also to identify and characterize low-energy alternative structures;
though they are not manifest in the bulk,
they can become physically relevant with
changes in the external parameters and boundary conditions produced in a nanocomposite.

In this paper, we use first principles calculations
to explore the structure and properties of 
a prototypical atomic-scale checkerboard of BiFeO$_3$ and BiMnO$_3$
(Fig~\ref{checkerboardfig}), extending
a shorter study of this nanocomposite that has been published 
elsewhere~\cite{Palova09short}.
Ferroelectric antiferromagnetic (AFM) bulk BiFeO$_3$ and half-metallic ferromagnetic (FM) bulk BiMnO$_3$
are good candidates for a nanocomposite with ferroelectric-ferromagnetic (multiferroic) behavior.
The properties of the atomic-scale checkerboard are found to be directly related
to the properties of the bulk constituents in their ground states and in low-energy alternative structures.
The ground state of the BiFeO$_3$-BiMnO$_3$ atomic-scale checkerboard 
is multiferroic, i.e. 
ferroelectric and ferrimagnetic, 
acquiring the desired properties from the constituents. 
In addition, we show that the BiFeO$_3$-BiMnO$_3$ atomic scale checkerboard displays a 
magnetostructural effect,
namely, it changes its magnetic ordering with the
change of its crystal structure.
This effect is argued to be inherent to B-site cation checkerboard geometry,
resulting from magnetic frustration for the particular arrangement of cations and bonds.

The organization of this paper follows.
In Sec.~\ref{methodsec}, we describe 
the first-principles method
and the 
structural distortions
and magnetic orderings considered.
Results for low-energy alternative structures of bulk 
BiFeO$_3$ and BiMnO$_3$ are reported 
in Secs.~\ref{bfosec} and~\ref{bmosec}, respectively.
The ground state of the BiFeO$_3$-BiMnO$_3$ atomic-scale checkerboard
is shown to be ferroelectric and ferrimagnetic
in Sec.~\ref{bfmogsstrucsec}.
A simple Heisenberg model is constructured to represent the energies of various magnetic states of this checkerboard computed from first principles.
In Sec.~\ref{bfmoaltsec}, 
the effect of structural distortions
on the magnetic ordering of the nanocheckerboard is explored,
and we relate the properties of alternative low-energy
structures of the checkerboard
to those 
of bulk BiFeO$_3$ and BiMnO$_3$.
Anisotropic epitaxial strain is shown to drive a magnetic transition 
in the atomic-scale checkerboard in Sec.~\ref{bfmostrainsec}.
We study the effect of B-site cation arrangement
on magnetic properties of the BiFeO$_3$-BiMnO$_3$ system
in Sec.~\ref{bfmomagidealsec}.
The possibility of
experimentally realizing a BiFeO$_3$-BiMnO$_3$ nanocheckerboard
is discussed in Sec.~\ref{discussionsec}.
Conclusions are presented 
in Sec.~\ref{summarysec}.

\section{Method}
\label{methodsec}

First-principles calculations are performed using density functional theory within the 
local spin-density approximation with Hubbard U
(LSDA+U) method as implemented in 
the Vienna 
{\sl ab initio} simulation package VASP-4.6.34~\cite{Kresse93,Kresse96}.
Projector-augmented wave potentials (PAW)~\cite{Blochl94,Kresse99}
are used with 15 valence electrons for Bi ($5d^{10}6s^2 6p^3$),
14 for Fe ($3p^6 3d^6 4s^2$), 13 for Mn ($3p^6 3d^5 4s^2$),
and 6 for O ($2s^2 2p^4$).
The robustness of the results is tested with two
different implementations of the rotationally invariant LSDA+U
version.
The first is due to Liechtenstein~\cite{Liechtenstein95}
with effective on-site Coulomb interaction $U_{Fe}=U_{Mn}=5 eV$
and effective on-site exchange interaction $J_{Fe}=J_{Mn}=1 eV$.
The second is due to Dudarev~\cite{Dudarev98},
with $U_{Mn}^{eff}=5.2 eV$, $U_{Fe}^{eff}=4 eV$, where $U^{eff}=U-J$.
Both implementations treat localized $d$ electron states in Fe and Mn.
It has previously been shown that these $U$ and $J$ values 
give good agreement with experiment in bulk BiFeO$_3$~\cite{Neaton05}.
The value $U_{Mn}^{eff}=5.2 eV$ has previously been used 
for bulk BiMnO$_3$ ground state calculations~\cite{Baettig07}.

\begin{figure} [h]
\begin{center}
\includegraphics[scale=0.45]{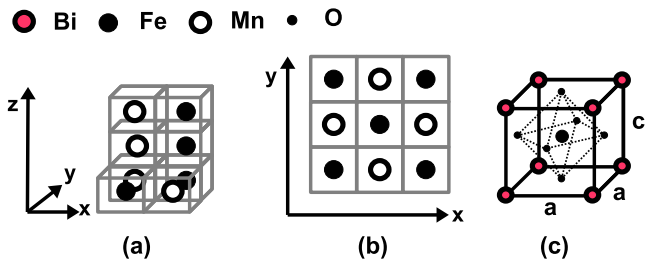}
\caption
        {
(a) BiFeO$_3$-BiMnO$_3$ atomic-scale checkerboard.
(b) Top view of the atomic-scale checkerboard.
	(c)   Perovskite cell.
	Dashed lines show an oxygen octahedron surrounding the 
	B-site (Fe, or Mn) cation.
	}
\label{checkerboardfig}
\end{center}
\end{figure}

The BiFeO$_3$-BiMnO$_3$ atomic-scale checkerboard is 
shown in Fig.~\ref{checkerboardfig}.
BiFeO$_3$ and BiMnO$_3$ alternate at the atomic level,
forming a checkerboard pattern in 
the xy plane and pillars of the same composition along z.
The supercell is $\sqrt{2}a \times \sqrt{2}a \times 2 c$, containing 
two Fe and two Mn.
In the limit of the atomic-scale pillars considered here,
the checkerboard structure is
the same as that of a (110)-oriented superlattice.

We consider two additional types of B-site
 cation-ordered
BiFeO$_3$-BiMnO$_3$ systems:
a (001)-oriented layered superlattice,
with single unit-cell Fe and Mn layers alternating along $z$,
and 
a rocksalt structure,
with Fe and Mn alternating in every other unit cell
((111) superlattice in the atomic-scale limit 
considered here).
In both cases, the supercell is $\sqrt{2}a \times \sqrt{2}a \times 2c$.
For consistency, we take the supercell
for bulk BiFeO$_3$ and bulk BiMnO$_3$ calculations to be
$\sqrt{2}a \times \sqrt{2}a \times 2c$,
except for the R3c structure, where we use a 
$\sqrt{2}a \times \sqrt{2}a \times \sqrt{2}a$ supercell.

\begin{figure} [h]
\begin{center}
\includegraphics[scale=0.35]{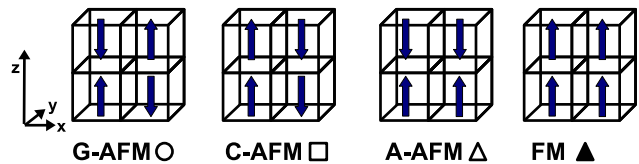}
\caption
{
Magnetic orderings considered for 
bulk BiFeO$_3$ and bulk BiMnO$_3$:
Symbols for each type of ordering are introduced next to each label.}
\label{CGtypeAFMFMfig}
\end{center}
\end{figure}

\begin{figure} [h]
\begin{center}
\includegraphics[scale=0.35]{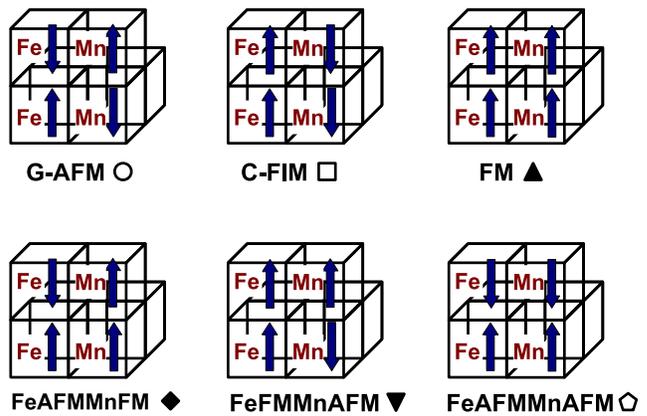}
\caption
{ 
Magnetic orderings considered for the BiFeO$_3$-BiMnO$_3$ atomic-scale checkerboard compatible with the $\sqrt{2}a \times \sqrt{2}a \times 2c$ supercell.
Symbols for each type of ordering are introduced next to each label.
}
\label{AFMFMnanofig}
\end{center}
\end{figure}

Several types of magnetic orderings
are studied here:  the G-type (rocksalt),
C-type,
A-type AFM,
and FM ordering 
of the local magnetic moments
in bulk BiFeO$_3$,
or bulk BiMnO$_3$ (see Fig.~\ref{CGtypeAFMFMfig}).
All orderings considered are collinear; this is supported by recent neutron scattering measurements
on BiFeO$_3$ doped with Mn~\cite{Sosnowska02},
that indicate
collinear AFM ordering.

Magnetic orderings
of the BiFeO$_3$-BiMnO$_3$ atomic-scale checkerboard are shown 
in Fig.~\ref{AFMFMnanofig},
where we consider six collinear orderings of Fe and Mn spins.
Similarly, six collinear 
orderings of the magnetic Fe and Mn spins are explored
in the BiFeO$_3$-BiMnO$_3$ (001)-oriented superlattice
and the rocksalt structure.
For the (001)-oriented superlattice,
these orderings are described by the
notation FeFM (FeAFM), or MnFM (MnAFM),
referring to the FM (AFM) ordering for the Fe (Mn) moments 
in the relevant layer, respectively,
with the remaining ambiguities resolved as follows: 
FeAFMMnAFM magnetic order has AFM ordered Fe and Mn layers with 
FM order along the mixed Fe-Mn chains in the $z$ direction,
while G-AFM designates the case with AFM order along the mixed chains; 
similarly, FeFMMnFM has FM ordered Fe and Mn layers with AFM order, 
while FM designates the case with FM order along the mixed
chains.
For the rocksalt structure,
we consider FM and G-AFM ordering,
FeAFMMnFM ordering,
referring to
FM ordered Mn sublattice and
AFM ordered Fe sublattice;
similarly we consider FeFMMnAFM ordering
with FM ordered Fe and AFM ordered Mn sublattices, respectively.
Finally, the FMFM ordering
has AFM ordered 
Mn and Fe sublattices, which are coupled FM in each Fe-Mn $z$ layer, 
while the AFMAFM ordering
has AFM ordered 
Mn and Fe sublattices coupled AFM in each Fe-Mn $z$ layer.

\begin{figure} [h]
\centering
\includegraphics[scale=0.33]{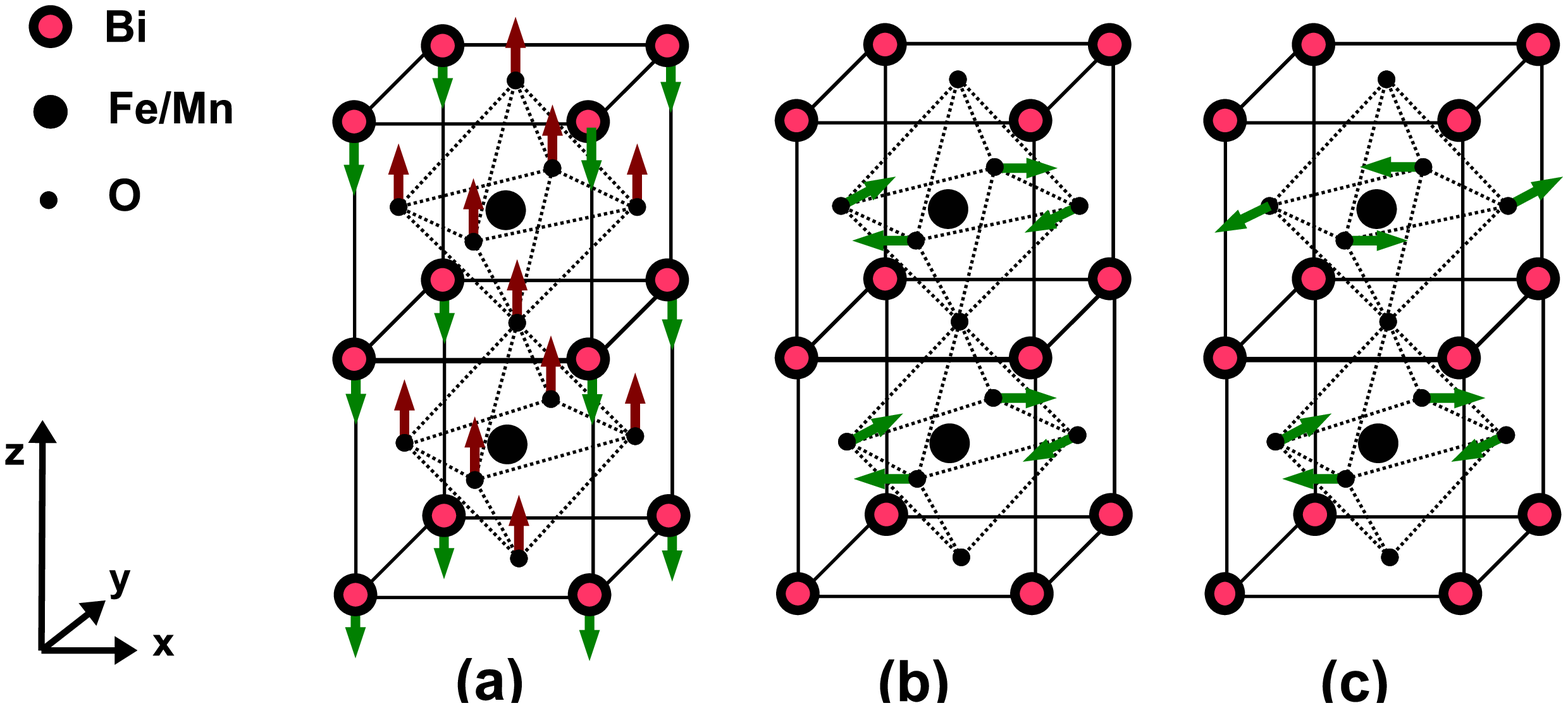}
\caption{
Distortions of the ideal cubic perovskite cell considered in this work~\cite{Stokes02}:
(a) Polar distortion with 
relative shift of Bi and Fe/Mn cations with respect to O 
anions along the $z$-axis ($\Gamma_4^-(z)$),
(b) $+$ (in-phase) rotations of the oxygen octahedra (dashed lines)about the $z$-axis ($M_3^+(z)$),
(c) $-$ (out-of-phase) rotations of the oxygen octahedra (dashed lines)
about the $z$-axis ($R_4^+(z)$).
}
\label{perovskitemodesfig} 
\end{figure}

\begin{table*}
\begin{center}
\caption{
Resulting space groups for considered structural distortions 
(see Fig.~\ref{perovskitemodesfig}).
Pm$\bar{3}$m is the ideal perovskite structure and
P4/mmm is the uniformly strained tetragonal unit cell.
} 
\label{structurestab}
{\small
\begin{tabular}{cccccccc}
\\
\hline
\hline
 Modes        & $\Gamma_4^-(z)$ &  $M_3^+(z)$ &  $M_3^+(z)$,$\Gamma_4^-(z)$  & $R_4^+(y)$  & $R_4^+(y)$,$\Gamma_4^-(y)$ & $R_4^+([111])$ & $R_4^+([111])$,$\Gamma_4^-([111])$\\
 Abbrevation & $\Gamma_4^-(z)$ &  $M_3^+(z)$ &  $M \Gamma (z)$  & $R_4^+(y)$  & $R \Gamma(y)$ & $R_4^+(d)$ & $R\Gamma(d)$\\
  Space Group & P4mm            &  P4/mbm      &  P4bm                        & I4/mcm       & I4cm                       & R$\bar{3}$c            & R3c \\
\hline
\end{tabular}
}
\end{center}
\end{table*}

Structures generated by three modes
of the cubic perovskite structure are considered
(see Fig.~\ref{perovskitemodesfig})~\cite{Stokes02}:
(i) the zone center polar $\Gamma_4^-$ mode, 
(ii) the $M_3^+$ 
 oxygen octahedron rotations 
 (all rotations about a given axis are in phase),
and
(iii) $R_4^+$ rotations
(sense of rotations alternates along the rotation axis).
Space groups corresponding to the structural distortions considered
are presented in Table~\ref{structurestab},
and we use the notation c-, l-, or r- to refer to the structural distortion of the 
B-site cation-ordered checkerboard, layered superlattice, or rocksalt
structure, respectively.
To search for the ground state for a given magnetic ordering and structure type,
we perform structural relaxation with the conjugate gradient algorithm.
Both the cell shape and the cell volume are relaxed; more specifically,
the ions are relaxed towards
equilibrium positions until the Hellmann-Feynman forces are less than $10^{-3} eV/\AA$.
An energy cutoff $550 eV$ for the plane wave basis set is used.
Convergence in the energy is reached
with precision $10^{-7} eV$.
A Monkhorst-Pack k-point grid~\cite{Monkhorst76} 
is generated with density 
$4\times4\times4$ for the
($\sqrt{2} \times \sqrt{2} \times \sqrt{2}$) supercell, 
and $4\times4\times2$ for 
the ($\sqrt{2}\times\sqrt{2}\times 2$)
supercell.
For magnetic energy calculations 
(Secs.~\ref{bfmomagr3csec} and~\ref{bfmomagidealsec}),
we use the energy cutoff $800 eV$, and
the Monkhorst-Pack k-point grid with density 
$6\times6\times4$.
Gaussian broadening
of the partial occupancies for each wavefunction is 
$0.05 eV$.
A tetrahedron method with Blochl corrections~\cite{Blochl94b}
is used
for the density of states (DOS) calculations,
with the Monkhorst-Pack k-point grid
$4\times4\times4$ for 
the ($\sqrt{2} \times \sqrt{2} \times \sqrt{2}$)
and $8\times8\times4$ for
the ($\sqrt{2}\times\sqrt{2}\times 2$)
supercell.

The rotational distortion can be quantified
using the oxygen octahedron rotational angle $\Theta$
defined specifically for each oxygen in the octahedron
as 
\begin{equation}
cos \Theta = \frac{\vec{u}\cdot \vec{v}}{|\vec{u}||\vec{v}|},
\end{equation}
where $\vec{u}$ 
is the shortest vector from the rotation axis to the reference position of the oxygen, 
and $\vec{v}$ is the shortest vector from the rotation axis to the position of the oxygen in the distorted structure.
The rotation axis is 
[001] and [010] for the 
$M_3^+(z)$ and $R_4^+(y)$ distortions respectively,
and
the threefold axis (body diagonal of the cube or distorted cube) 
for the $R_4^+([111])$ distortion.
Due to deformation of the oxygen octahedron in the 
BiFeO$_3$-BiMnO$_3$ checkerboard structures,
these angles may be different for different oxygens in the same octahedron.
We report an average value if the range is small;
otherwise the lower and upper limits of the range are presented.

The polar distortions of the various structures
can be quantified by estimating the polarization based on a linearized expression with nominal charges: 
\begin{equation}
\vec{P} 
= \frac{|e|}{\Omega} \sum_{j} q_j \Delta \vec{u_j},
\label{pointchargeP}
\end{equation}
where $\vec{P}$ is the polarization,
$\Delta \vec{u_j}$ is the displacement of the $j$th ion with respect to its ideal perovskite position,
$q_j$ is the nominal charge of the $j$th ion
($q_{Bi}=+3$, $q_{Fe}=+3$, $q_{Mn}=+3$, $q_O=-2$),
and $\Omega$ is the unit cell volume.

For selected structures the true value for the spontaneous polarization 
is computed using the 
Berry phase method~\cite{King-Smith93,Resta07}
as implemented in VASP--4.6.34.
In this formalism,
the polarization is only well-defined {\it mod} $e\vec{R}/\Omega$,
where $\vec{R}$ is any lattice vector and $\Omega$ is the primitive-cell volume;
thus possible values of the polarization are points on the lattice defined by $\vec{P_0} + e\vec{R}/\Omega$, where $\vec{P_0}$ is the value directly obtained from the Berry phase calculation.
Choosing the lattice point (or "branch") that corresponds to the measured switching polarization (e.g. in an electrical hysteresis loop) 
is done by
computing the polarization of states closely spaced along an adiabatic path connecting the structure of interest to a high-symmetry reference structure.
These laborious calculations can be avoided by
an approach based on the reformulation of the polarization in terms of  Wannier function centers~\cite{King-Smith93};
the switching polarization is obtained from the difference between the two symmetry-related variants by associating the Wannier centers 
with the same atoms in both structures~\cite{Marzari97,OscarKarin}.
Due to incompatibility between the Wannier90 and the VASP
codes, we cannot use this latter approach here; 
we make the necessary branch choices based on computations along adiabatic paths combined with the nominal-charge polarization estimate.

\section{BiFeO$_3$ Structures}
\label{bfosec}

\begin{figure} [h]
\begin{center}
\includegraphics[scale=0.7]{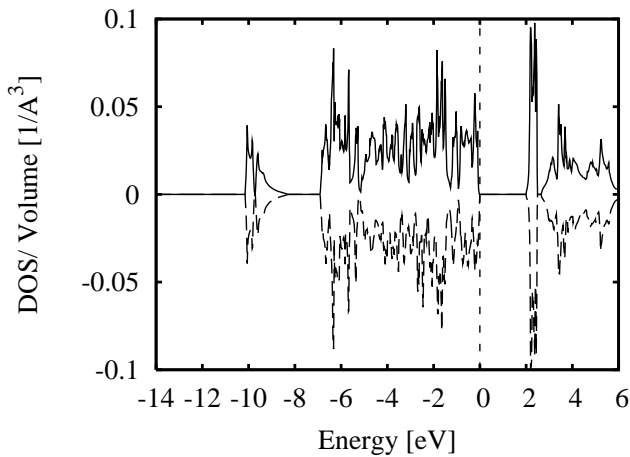}
\caption
{Density of states (DOS) of the R3c G-AFM ground state (GS) of BiFeO$_3$.
Spin up states are plotted by solid line and 
spin down states by a dashed line (inverted view).
The zero is set to the valence band maximum.
DOS shows an insulating band gap of 2 eV.}
\label{dosBFOfig}
\end{center}
\end{figure}

In agreement with previous first-principles calculations
and experiment~\cite{Neaton05,Michel69,Kubel90,Sosnowska02},
we find that the ground state structure of BiFeO$_3$
 has rhombohedral R3c symmetry, which is a combination of the rotational $R_4^+([111])$ mode (counter-rotations of the oxygen octahedra about the [111] axis) and a polar $\Gamma_4^-([111])$ 
modes, with Bi, Fe, and O displaced relative to one another along
[111] and further distortion of the oxygen octahedra by displacement of the O displaced perpendicular to [111].~\cite{Stokes02}
The oxygen octahedra rotation angle
is large,
about $14^{\circ}$,
and is comparable to rotations in other perovskites~\cite{Coh09}.
The R3c ground state has G-AFM (rocksalt) ordering
(see Fig.~\ref{CGtypeAFMFMfig}),
and Fe local magnetic moment of $4 \mu_B$.~\cite{Sosnowska02}
The density of states (DOS)  
is plotted in Fig.~\ref{dosBFOfig}:
it has a $2 eV$ band gap
that separates occupied and unoccupied Fe $d$ states.
The polar character of BiFeO$_3$ arises from the polar $\Gamma_4^-([111])$
mode,
and the spontaneous polarization
using the Berry phase method is
$P^{R3c} = 90 \mu C/cm^2$ along the [111] direction.

\begin{figure} [h]
\begin{center}
\includegraphics[scale=0.35,angle=-90]{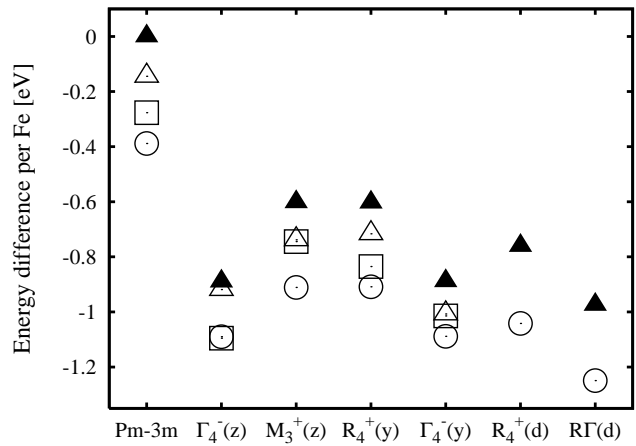}
\caption
{Structural energetics of bulk BiFeO$_3$. Energy difference per Fe
for different magnetic orderings (see Fig.~\ref{CGtypeAFMFMfig})
and structural distortions (see Fig.~\ref{perovskitemodesfig} and Table~\ref{structurestab}) 
relative to the FM Pm$\bar{3}$m structure.
}
\label{ediffdistbfofig}
\end{center}
\end{figure}

Next
alternative structures of BiFeO$_3$ are studied, and we consider those 
generated by freezing in linear
combinations of the
rotational $M_3^+$, $R_4^+$, and polar $\Gamma_4^-$ modes,
and four magnetic orderings
(see Fig.~\ref{CGtypeAFMFMfig}).
Their energies are plotted
in Fig.~\ref{ediffdistbfofig} 
(see also Table~\ref{BFOtab}).
By symmetry, the FM ordering has the same energy for the $\Gamma_4^-(z)$ and $\Gamma_4^-(y)$ structures; this is also true for the G-AFM ordering.
In contrast, the C-AFM ordering has different energy for the $\Gamma_4^-(z)$ and $\Gamma_4^-(y)$ structures; this is also true for the A-AFM ordering, as the y and z directions for these spin arrangements are not symmetry-related 
(see Figs.~\ref{CGtypeAFMFMfig} and~\ref{ediffdistbfofig}).

For all structural distortions considered, 
the favored magnetic ordering is G-AFM (open circle).
This is consistent with the Goodenough-Kanamori rules:
either a
strong $\sigma$ bond is formed between Fe $e_g$ and the neighbouring 
O $p$ orbitals
in an ideal $180^{\circ}$ Fe-O-Fe bond
(ideal perovskite structure),
or a weak $\pi$ bond is formed between Fe $t_{2g}$ and O $p$ orbitals 
when the bond is bent towards $90^{\circ}$
(as the structure is distorted);
in both cases, the AFM superexchange  
is favored~\cite{Goodenough55,Kanamori59,LeeAFM}.

The most favorable low-energy alternative structures and 
the ground state of
BiFeO$_3$ are presented in Table~\ref{BFOtab}.
The low-energy $\Gamma_4^-(z)$ structure, with P4mm symmetry, is the "supertetragonal"
structure
with $c/a \sim 1.3$, previously discussed
elsewhere~\cite{Ederer05tetrag,Hatt10}.
It has been shown recently that
this phase can be stabilized 
in BiFeO$_3$ thin films~\cite{Zeches09}.
The polarization computed by the Berry phase method~\cite{Ederer05tetrag} 
is $P^{P4mm}\approx 150 \mu C/cm^2$.
The nominal-charge estimates are therefore smaller than the true values for both 
the P4mm and R3c structures, but the relative 
values are well reproduced. The polar distortion is dominant in all structures considered; the 
rotation-only structures are higher in energy, and the presence of a polar distortion tends to stabilize the rotational instabilities.
For example,
the $\Gamma_4^- $+$ M_3^+$ and $\Gamma_4^- $+$ R_4^+$ distortions
relax back to $\Gamma_4^-$ with zero 
oxygen octahedron rotation angle.
The only exception is the R3c structure, in which the rotational and polar distortions coexist. 

\begin{table}
\begin{center}
\caption{
 GS and low-energy alternative structures 
 of bulk BiFeO$_3$.
 The energy difference $\Delta E$ is given with respect to
 the FM Pm$\bar{3}$m structure, as in Fig.~\ref{ediffdistbfofig}.
 Polarization $P$ is
 estimated from the nominal charges (Eq. (\ref{pointchargeP})).
 Also included are the band gap $\Delta$ and the a and c lattice constants for the 
 $\sqrt{2}a \times \sqrt{2}a \times 2c$ supercell of P4mm,
 and the a lattice constant and the angle $\alpha$  
 for the $\sqrt{2}a \times \sqrt{2}a \times \sqrt{2}a$ supercell of R3c.
} 
\label{BFOtab}
{\small
\begin{tabular}{cccc}
\\
\hline
\hline
 Space group     & P4mm            &             & R3c \\
 Modes           & $\Gamma_4^- (z)$ &             & $R_4^+([111])$, $\Gamma_4^-([111])$\\
 \hline
 Mag. order      & G-AFM           &C-AFM        & G-AFM  \\
$\Delta E$[eV/Fe]&-1.09            &-1.10        &-1.25 \\
 $\Delta$ [eV]   &1.75             &2.23         &1.99 \\
 P[$\mu C/cm^2$] &113.6            &116.2        & 62.1 \\
 a/\,c [\AA]     &3.68/\,4.64      &3.67/\,4.68  & 5.52,\,$\alpha=59.8^{\circ}$ \\
\hline
\hline
\end{tabular}
}
\end{center}
\end{table}

\section{BiMnO$_3$ Structures}
\label{bmosec}

Previous first principles calculations show that the ground state structure
of bulk BiMnO$_3$
is monoclinic centrosymmetric C2/c
with zero spontaneous polarization~\cite{Baettig07} and FM ordering
~\cite{dosSantos02b,Atou99,Shishidou04}.
Bismuth (Bi) cations are off-center due to stereochemically active Bi lone pairs,
and the Jahn-Teller activity of Mn$^{3+}$
further distorts the structure~\cite{Seshadri01}.
Optimizing the atomic positions and lattice constants, 
we performed a first-principles calculation for this structure 
to find an energy gain of 
$1.26 eV$/Mn relative to the the ideal cubic perovskite structure with G-AFM ordering and 
$a_0 = 3.83 \AA$; the latter is used as our reference state throughout this section. 

\begin{figure} 
\begin{center}
\includegraphics[scale=0.7]{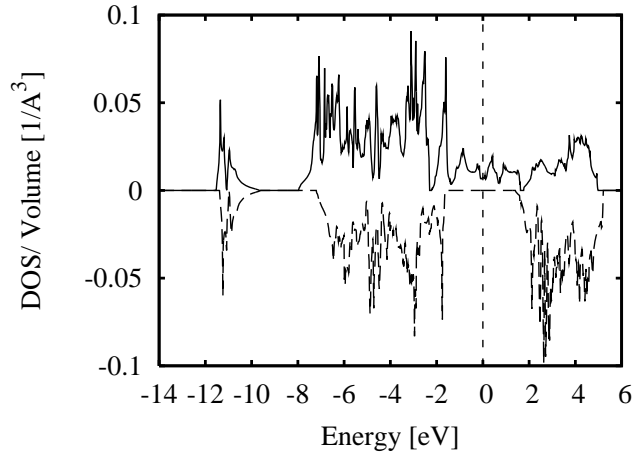}
\caption
{Density of states (DOS)
of the R3c FM alternative structure of BiMnO$_3$.
Spin up states are plotted by solid line and 
spin down states by a dashed line (inverted view).
Fermi energy is shown by the vertical dashed line crossing zero.
DOS shows a half-metallic
character.}
\label{dosBMOfig}
\end{center}
\end{figure}

\begin{figure} [h]
\begin{center}
\includegraphics[scale=0.35,angle=-90]{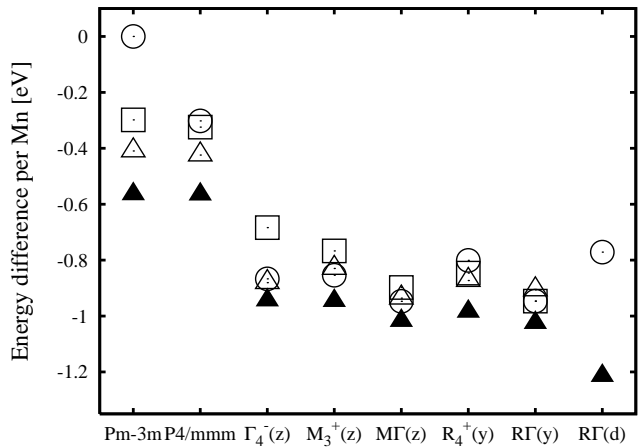}
\caption
{
Structural energetics of bulk BiMnO$_3$. Energy difference per perovskite cell 
(Mn) 
for different magnetic orderings 
(see Fig.~\ref{CGtypeAFMFMfig})
and for structural distortions (see Fig.~\ref{perovskitemodesfig} and Table~\ref{structurestab}) 
generated by the specified modes. 
}
\label{ediffdistbmofig}
\end{center}
\end{figure}

We study low-energy alternative structures of BiMnO$_3$ compatible with a 
$\sqrt{2}\times \sqrt{2}\times 2$ supercell.
Results are presented in Fig.~\ref{ediffdistbmofig}.
The lowest energy structure 
has R3c symmetry,
the same structure type as the ground state of BiFeO$_3$.
It is FM, with magnetic moment $3.9\mu_B$ per Mn.
This structure lies  
only $43 meV$/Mn above the BiMnO$_3$ monoclinic ground state.
The computed DOS is shown in Fig.~\ref{dosBMOfig}:
the system is half-metallic,
with a gap of 3.25 eV in the spin down channel. As an aside, we note that
it might be useful to stabilize BiMnO$_3$ as a half-metal in this low energy structure
for possible applications in spintronics~\cite{Wolf01}.

For all structural distortions considered, 
the favored magnetic ordering is FM, consistent with previous analysis that showed that
BiMnO$_3$ favors FM structures with a half-metallic character~\cite{Seshadri01,Hill99}.
The ferromagnetism in BiMnO$_3$ can be explained by a combination
of Goodenough-Kanamori rules and
orbital ordering~\cite{dosSantos02b,Yang06,Hill00,Seshadri01}.
Structural distortions
(either oxygen octahedron rotations or polar distortion)
widen the spin-down gap
(see Table~\ref{BMOtab});
a similar trend is observed for the band gap in BiFeO$_3$
(see Table~\ref{BFOtab}).
A small band gap opens with a monoclinic distortion 
in the FM BiMnO$_3$ ground state~\cite{Seshadri01,Shishidou04}.

\begin{table*}
\begin{center}
\caption{
 Low energy alternative structures
 of FM bulk BiMnO$_3$.
 The energy difference $\Delta E$ is calculated
 with respect to the G-AFM Pm$\bar{3}$m structure
 (as in Fig.~\ref{ediffdistbmofig}).
 Listed are values of 
 the spin-down band gap $\Delta_{hm}$ in the half-metallic structures
 or metallic (m) character,
 the oxygen octahedron rotational angle $\Theta$
 (see Sec.~\ref{methodsec}),
 and the 
 $a$ and $c$ lattice constants of the $\sqrt{2}\times \sqrt{2} \times 2$ supercell.
 } 
\label{BMOtab}
{\small
\begin{tabular}{ccccc}
\\
\hline
\hline
 Space group          & P4/mmmm   & P4bm      & I4cm      & R3c   \\
 Modes                & -         &$M_3^+(z)$, $\Gamma_4^-(z)$ & $R_4^+(y)$, $\Gamma_4^-(y)$ &  $R_4^+([111])$, $\Gamma_4^-([111])$ \\
 \hline
 $\Delta E$ [eV/Mn] &-0.57      &-1.02      &-1.03      &-1.22  \\
 $\Delta_{hm}$ [eV]    &  m        &0.73       &2.74      & 3.25   \\
 $\Theta  [^{\circ}]$  &-          &11.4       &12.0       &13.3 \\
 a/ c [\AA]           &3.83/ 3.86 &3.81/ 4.01 &3.81/ 3.83 &5.51, $\alpha=60^{\circ}$ \\
\hline
\hline
\end{tabular}
}
\end{center}
\end{table*}

The Jahn-Teller active Mn$^{3+}$ configuration tends to favor elongation of the oxygen octahedron.
In contrast to BiFeO$_3$, in which the polar instability strongly dominates, the rotational and polar instabilities in BiMnO$_3$ are comparable in magnitude, as can be seen by comparing the energies of
 the $\Gamma_4^-(z)$, $M_3^+(z)$ and $R_4^+(y)$ states. The latter two states have a small residual polar instability. 
A polar distortion along a Cartesian axis lowers the energy of the G-AFM state so that the energy difference between this state and the FM ground state is greatly reduced; this does not occur if the polar distortion is along [111] as in the R3c phase. 
The octahedral rotation angles in the low-energy BiMnO$_3$ structures are all similar in magnitude, varying between 
$11 - 14^{\circ}$,
with an angle of $13^{\circ}$ for the FM R3c structure.
The value of the octahedral rotation angle in the G-AFM R3c structure, $14^{\circ}$, is the same as in G-AFM R3c BiFeO$_3$.

\section{BiFeO$_3$-BiMnO$_3$ Nanocheckerboard Ground State}
\label{bfmogsstrucsec}

\subsection{Crystal structure, magnetization and polarization}
\label{bfmogssec}

In the search for the ground state of the atomic-scale checkerboard cation
ordering, we considered the six collinear magnetic states of 
Fig.~\ref{AFMFMnanofig} and four different structures:
the tetragonal P4/mmm structure and three additional structures, 
obtained by freezing in a $\Gamma_4^-$(z) 
mode, a combination of R$_4^+$(y) and $\Gamma_4^-$(y), 
and a combination of R$_4^+$(111) and $\Gamma_4^-$(111).
We designate these latter three structures by the space group they 
would have if all B sites were occupied by the same cation, 
with the prefix c- to remind us that the actual symmetry is 
lower due to the checkerboard ordering: c-P4mm, c-I4cm, and c-R3c. 
The GS 
of the BiFeO$_3$-BiMnO$_3$ nanocheckerboard
is found to be c-R3c, as could be expected based on the R3c GS of bulk BiFeO$_3$,
and on our results for bulk BiMnO$_3$. 
The magnetic ordering in the c-R3c GS is FeAFMMnFM.
Fe magnetic moments are ordered AFM
along the Fe pillars and
Mn magnetic moments are ordered FM along the Mn pillars,
as expected from the G-AFM and FM ground states of BiFeO$_3$ and BiMnO$_3$, respectively
(see Secs.~\ref{bfosec} and~\ref{bmosec}).
AFM and FM $xy$ layers alternate along $z$ as is sketched in
Fig.~\ref{AFMFMnanofig}.
The computed Fe and Mn 
local magnetic moments
are $4.1 \mu_B$ and $3.8 \mu_B$, respectively;
these are the same values as those reported here in the parent compounds BiFeO$_3$ and BiMnO$_3$.
Although the contribution from Fe magnetic moments to the net magnetization 
cancels due to the AFM pillar ordering,
the contribution from Mn moments adds,
leading to a net magnetization of $3.8 \mu_B$  
per Fe-Mn pair.

\begin{figure} [h]
\begin{center}
\includegraphics[scale=0.7]{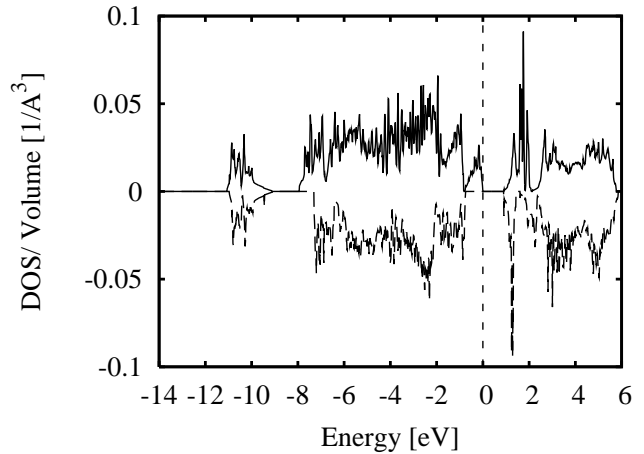}
\caption
{Density of states (DOS)
of the c-R3c FeAFMMnFM ground state (GS) of BiFeO$_3$-BiMnO$_3$ nanocheckerboard. 
Spin up states are shown by a solid line and spin down states by a dashed line.
The zero is set to the valence band maximum.
The band gap in the spin-up channel is 0.9 eV.}
\label{dosBFMOfig}
\end{center}
\end{figure}

The DOS of the c-R3c FeAFMMnFM GS
is shown in Fig.~\ref{dosBFMOfig}.
The general features are very similar to those found in BiFeO$_3$ 
(Fig.~\ref{dosBFOfig}) and BiMnO$_3$ (Fig.~\ref{dosBMOfig}), 
the main difference being that the spin-up Mn states at the 
Fermi level in BiMnO$_3$ have split to open a gap, 
with the occupied states at the top of the valence band narrowing
the gap to $0.9 eV$.

\begin{figure} [h]
\begin{center}
\includegraphics[scale=0.35,angle=-90]{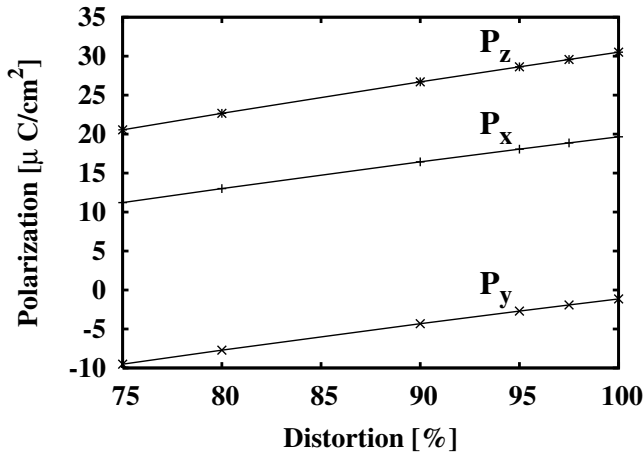}
\caption
{
Spontaneous polarization $\vec{P} = (P_x,P_y,P_z)$ as a function of the 
structural distortion of the c-R3c FeAFMMnFM BiFeO$_3$-BiMnO$_3$ 
nanocheckerboard.
$100\%$ distortion corresponds to the c-R3c ground state structure, 
and $0\%$ to the ideal perovskite structure.
}
\label{totpolarfig}
\end{center}
\end{figure}

Direct calculation of the spontaneous polarization using the Berry phase method for the c-R3c GS yields a value of
$\vec{P} = (19.6,\; -1.1,\; 30.5)$ $\mu C/cm^2$.
This is well defined only up to the polarization lattice
$e\vec{R}/\Omega$~\cite{King-Smith93,Resta07}, 
which in this case is
$(13.2,\; 13.1,\; 0.1)n_1 + 
(-13.2,\; 13.0,\; 0.1)n_2 +
(0.0,\; 0.2,\; 26.6)n_3 \; \mu C/cm^2$,
where $\vec{n}$ is a vector of integers.
To determine the branch that corresponds to the switching polarization, we compute the polarization along a structural deformation path that linearly connects the c-R3c GS to the ideal cubic perovskite structure.
As shown in Fig.~\ref{totpolarfig}, 
the computation is performed for structures down to 75\% of the full distortion (at which point the structures 
become metallic) and then linearly extrapolated to 0\% using the expression 
\begin{equation}
\Delta \vec{P}_{100\%-0\%} = 4 \times \Delta \vec{P}_{100\%-75\%} =
(33.8, 33.5, 39.8) \mu C/cm^2.
\label{bfmoberryphase}
\end{equation}
The magnitude of this estimate, $62.0 \mu C/cm^2$, 
suggests the branch choice
$\vec{P}_{bp} = (32.9,\; 38.0,\; 30.7)$ $\mu C/cm^2$
with magnitude $|P_{bp}|=58.9 \mu C/cm^2$.
However, we would expect it to be considerably larger,
based on comparison between the Berry phase $|P|$
of bulk BiFeO$_3$ and the $|P|$
computed using nominal charges (see Sec.~\ref{bfosec});
following this intuition we would make the branch choice
$\vec{P}_{bp} = (46.1,\; 51.3,\; 57.4)$ $\mu C/cm^2$
with magnitude $|P_{bp}|=89.7 \mu C/cm^2$.
This remaining ambiguity highlights the challenge of picking the right branch when the polarization is much larger than the quantum; in either case it is clear that the polarization of the checkerboard is comparable to the largest values found in ferroelectrics.
Thus, we find that the c-R3c GS of the
 BiFeO$_3$-BiMnO$_3$ nanocheckerboard
is multiferroic:
ferroelectric, with polarization
comparable to the polarization of bulk BiFeO$_3$,
and ferrimagnetic, with magnetization contributed by ferromagnetic
ordering in the BiMnO$_3$ component.

\subsection{Magnetic Coupling Constants}
\label{bfmomagr3csec}

\begin{figure} [h]
\begin{center}
\includegraphics[scale=0.65]{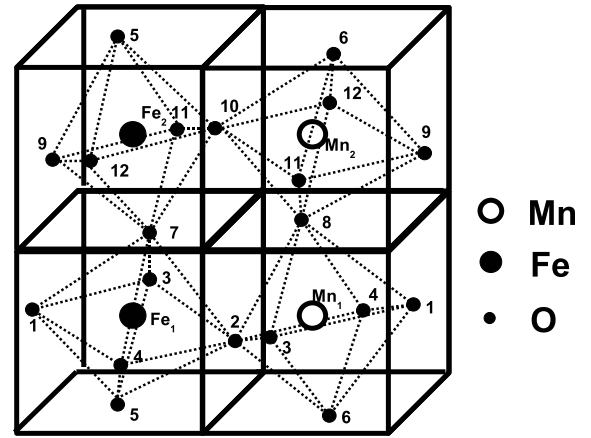}
\caption
{ 
Sketch showing the displacements of oxygen atoms in the 
$R_4^+([111])$ mode,
which contributes to the c-R3c
ground state (GS) of the BiFeO$_3$-BiMnO$_3$ nanocheckerboard.
Two inequivalent iron (Fe$_1$ and Fe$_2$) and
manganese (Mn$_1$ and Mn$_2$) atoms
and
twelve oxygens 
($1-12$) comprise the $\sqrt{2}\times\sqrt{2}\times 2$ unit cell.
The corners of each cube are occupied by Bi (not shown).
}
\label{groundstateposfig}
\end{center}
\end{figure}

\begin{table*}
\begin{center}
\caption{
B-site-cation-oxygen-B-site-cation (B-O-B) bonds
in the c-R3c ground state (GS)
of the BiFeO$_3$-BiMnO$_3$ nanocheckerboard.
B-O and O-B bond lengths and the B-O-B bond angle are given.
Atoms are numbered as in Fig.~\ref{groundstateposfig}.
The subscript indicates the cartesian direction along which the bond lies.
} 
\label{bondstab}
{\small
\begin{tabular}{ccccc}
\\
\hline
\hline
 B-O-B bond          & Notation & 
 $|B-O|$ [\AA] & $|O-B|$ [\AA] & Angle \\
\hline
 $(Fe_1-O_7-Fe_2)_z$ & $J_{Fe}$ & 1.93 & 2.08 & 153.8$^\circ$ \\  
 $(Fe_2-O_5-Fe_1)_z$ & $J_{Fe}$ & 1.96 & 2.08 & 156.8$^\circ$  \\
 $(Mn_1-O_8-Mn_2)_z$ & $J_{Mn}$ & 2.10 & 1.91 & 153.7$^\circ$  \\
 $(Mn_2-O_6-Mn_1)_z$ & $J_{Mn}$ & 1.87 & 2.18 & 156.3$^\circ$  \\
 $(Mn_1-O_1-Fe_1)_x$ & $J_{int}^\alpha$  & 1.92 & 2.06 & 166.8$^\circ$  \\
 $(Fe_1-O_2-Mn_1)_x$ & $J_{int}^\beta$  & 1.97 & 1.97 & 156.8$^\circ$  \\
 $(Mn_1-O_3-Fe_1)_y$ & $J_{int}^\alpha$  & 1.95 & 1.95 & 165.1$^\circ$  \\
 $(Fe_1-O_4-Mn_1)_y$ & $J_{int}^\beta$  & 2.04 & 1.91 & 155.3$^\circ$  \\
 $(Mn_2-O_9-Fe_2)_x$ & $J_{int}^\beta$  & 2.05 & 2.03 & 152.0$^\circ$  \\
 $(Fe_2-O_{10}-Mn_2)_x$ & $J_{int}^\gamma$ & 1.94 & 2.11 & 143.3$^\circ$  \\
 $(Mn_2-O_{11}-Fe_2)_y$ & $J_{int}^\beta$ & 2.05 & 1.95 & 151.1$^\circ$  \\
 $(Fe_2-O_{12}-Mn_2)_y$ & $J_{int}^\gamma$ & 2.07 & 1.98 & 144.1$^\circ$  \\
\hline
\hline
\end{tabular}
}
\end{center}
\end{table*}

To gain insight into the magnetic properties of the nanocheckerboard, 
we model the magnetic ordering energies using a nearest-neighbor (nn) 
Heisenberg model. 
The nn magnetic couplings arise from superexchange 
through the oxygens that lie on the bonds between the B site cations, 
with the strength of the superexchange being quite sensitive to the 
geometry of the B-O-B' bond.
If the structure were ideal cubic perovskite, there would be three
independent couplings, J$_{Fe}$, J$_{Mn}$ and J$_{int}$, corresponding to
180$^\circ$ Fe-O-Fe, Mn-O-Mn and Fe-O-Mn bonds, respectively.
The analysis of the couplings in the c-R3c structure is based on the geometry of the B-O-B' bonds as given in Table ~\ref{bondstab}; the labeling of the bonds and the changes in the bonds due to the R$_4^+$([111]) mode are 
shown in Fig.~\ref{groundstateposfig}.
The two Fe-O-Fe bonds are
almost identical in bond angle and bond length; this is also the case for the two Mn-O-Mn bonds.
This suggests that a single value of $J_{Fe}$ and $J_{Mn}$ 
can be used for the Fe-O-Fe and Mn-O-Mn interactions, respectively.
On the other hand,
the mixed Fe-O-Mn bonds vary in both B-O bond length, from 
$1.91 - 2.11 \AA$,
and B-O-B bond angle, from $143.3 - 166.8^\circ$.
This suggests the use of three different coupling constants 
$J_{int}^{\alpha}$, $J_{int}^{\beta}$ or $J_{int}^\gamma$ 
for the Fe-O-Mn interactions based on the typical values of the bond angles, approximately 166$^\circ$, 154$^\circ$ and 144$^\circ$, respectively.
Note that the angles of the Fe-O-Mn bonds in the Fe$_1$-Mn$_1$ 
layer in Fig.~\ref{groundstateposfig} 
are about 166$^\circ$ and 154$^\circ$,
while the angles in the Fe$_2$-Mn$_2$ layer
are about 152$^\circ$ and 144$^\circ$.

\begin{table*}
\begin{center}
\caption{
Calculated magnetic energies $\Delta E$ 
per four-perovskite unit cell (u.c.)
in the c-R3c GS structure
of BiFeO$_3$-BiMnO$_3$ nanocheckerboard.
The notation for magnetic ordering is that of Fig.~\ref{AFMFMnanofig}.
The symbols $x,\,y,\,a,\,b,\,c$ appearing in the magnetic energy are defined as follows:
$x= J_{Fe} S_{Fe}S_{Fe}$, $y= J_{Mn} S_{Mn}S_{Mn}$,
$a= J_{int}^\alpha S_{Fe} S_{Mn}$,
$b= J_{int}^\beta S_{Fe} S_{Mn}$,
$c= J_{int}^\gamma S_{Fe} S_{Mn}$.
} 
\label{magenergytab}
{\small
\begin{tabular}{ccccc}
\\
\hline
\hline
 Magnetic state       & Magnetic ordering & Heisenberg energy & $\Delta$ E & Fitted $\Delta$ E \\
       &  & [per u.c.] & [eV/u.c.] &  [eV/u.c.] \\
\hline
 FeAFMMnFM & Fe$_1^\uparrow$ Fe$_2^\downarrow$ Mn$_1^\uparrow$ Mn$_2^\uparrow$   &
 $ E_0 -2x + 2y + 2a - 2c $ & 
 0.000 & 0.000\\
 FeAFMMnFM & Fe$_1^\downarrow$ Fe$_2^\uparrow$ Mn$_1^\uparrow$ Mn$_2^\uparrow$   & 
$ E_0 -2x + 2y -2a +2c $ & 
0.200 & 0.207\\
 G-AFM & Fe$_1^\uparrow$ Fe$_2^\downarrow$ Mn$_1^\downarrow$ Mn$_2^\uparrow$      &
 $ E_0 -2x -2y -2a -4b -2c$ & 
0.032 & 0.026\\
 C-AFM & Fe$_1^\uparrow$ Fe$_2^\uparrow$ Mn$_1^\downarrow$ Mn$_2^\downarrow$      & 
$ E_0 + 2x + 2y -2a -4b -2c$ & 
0.143 & 0.152\\
 FeFMMnAFM & Fe$_1^\uparrow$ Fe$_2^\uparrow$ Mn$_1^\downarrow$ Mn$_2^\uparrow$   & 
$ E_0 + 2x -2y -2a +2c$ & 
0.436 & 0.436\\
 FeFMMnAFM & Fe$_1^\uparrow$ Fe$_2^\uparrow$ Mn$_1^\uparrow$ Mn$_2^\downarrow$   &
$ E_0 + 2x -2y +2a -2c$ & 
0.222 & 0.229\\
 FeAFMMnAFM & Fe$_1^\uparrow$ Fe$_2^\downarrow$ Mn$_1^\uparrow$ Mn$_2^\downarrow$ & 
$ E_0 -2x -2y + 2a +4b + 2c$  & 
0.275 & 0.284\\
 FM & Fe$_1^\uparrow$ Fe$_2^\uparrow$ Mn$_1^\uparrow$ Mn$_2^\uparrow$            &
$ E_0 + 2x +2y + 2a + 4b + 2c$ & 
0.416 & 0.410\\
\hline
\hline
\end{tabular}
}
\end{center}
\end{table*}

The values of these five exchange couplings were determined from first-principles results for the total energies of various magnetic orderings for the c-R3c GS structure
of the nanocheckerboard, given in Table~\ref{magenergytab}.
The structure is fixed to that obtained for the
FeAFMMnFM ordering
Fe$_1^\uparrow$ Fe$_2^\downarrow$ Mn$_1^\uparrow$ Mn$_2^\uparrow$.
The ordering
Fe$_1^\downarrow$ Fe$_2^\uparrow$ Mn$_1^\uparrow$ Mn$_2^\uparrow$, also described as FeAFMMnFM, is a distinct state with a different (higher) energy. 
Similarly, for the FeFMMnAFM ordering, there are two distinct magnetic states:
Fe$_1^\uparrow$ Fe$_2^\uparrow$ Mn$_1^\downarrow$ Mn$_2^\uparrow$ and
Fe$_1^\uparrow$ Fe$_2^\uparrow$ Mn$_1^\uparrow$ Mn$_2^\downarrow$,
with different energies as given in Table~\ref{magenergytab}.

We express the 
Heisenberg magnetic energy of each magnetic state,
\begin{equation}
\label{magneticE}
E = E_0 + \frac{1}{2} \sum_{ij} J_{ij} S_i S_j,
\end{equation}
where  $S_i$ and $S_j$ are the spins
$S_{Fe}=\frac{5}{2}, S_{Mn}=\frac{4}{2}$
with coupling constants 
$J_{ij} = J_{Fe}$, $J_{Mn}$, $J_{int}^\alpha$, $J_{int}^\beta$, $J_{int}^\gamma$,
and $E_0$ is a constant.
We extract values of the coupling constants 
by fitting the Heisenberg model energy to 
the first-principles energies by the least-squares method, obtaining
\begin{eqnarray}
\label{Jeffvalues}
& & E_0 = 218 meV, \quad J_{Fe} = 7.1 meV , \quad
J_{Mn} = -3.2 meV , \cr
& &
J_{int}^\alpha = -3.0 meV , \quad
J_{int}^\beta = 4.3 meV , \quad
J_{int}^\gamma = 7.3 meV. \nonumber
\\
\end{eqnarray}
The quality of the fit can be assessed by comparing the first-principles
energy to the fitted values in the fifth column of the table.

The AFM character of $J_{Fe}$
and the FM character of $J_{Mn}$ 
correspond to that of
bulk G-AFM BiFeO$_3$ and
bulk FM BiMnO$_3$, respectively.
Their values are comparable to those obtained from the observed bulk transition
temperatures within mean field theory assuming a single J: 
$ J_{Fe,bulk}\approx 6.3 meV$ and
$ J_{Mn,bulk}\approx -1.5 meV$,
respectively~\cite{Kiselev63,Chiba97,Sugawara65,Sugawara68,Faqir99,Baettig05prb}.
The correspondence is not exact because of the difference in the bond geometry between  
bulk BiFeO$_3$ and bulk BiMnO$_3$ and the nanocheckerboard.

\begin{figure} [h]
\begin{center}
\includegraphics[scale=0.5]{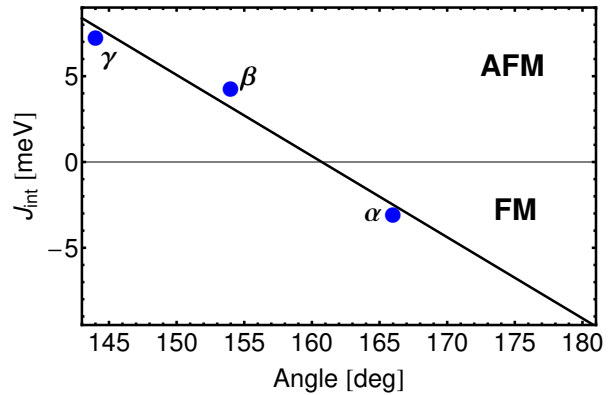}
\caption
{ 
Angular dependence of the Fe-Mn magnetic exchange
coupling constant $J_{int}$
in the c-R3c
GS structure of the BiFeO$_3$-BiMnO$_3$ nanocheckerboard (blue circles).
The black line is a linear fit. 
$J_{int}$ changes sign
(AFM $J_{int}>0$ to FM $J_{int}<0$) at the bond angle of $160^\circ$.
}
\label{Jintanglefig}
\end{center}
\end{figure}

In the Fe$_1$-Mn$_1$ layer,
the average Fe-Mn interaction $J_{int}$ is very weakly AFM
$(J_{int}^\alpha+J_{int}^\beta)/2 \gtrsim 0$, while
in the Fe$_2$-Mn$_2$ layer
it is strongly AFM 
$(J_{int}^\beta+J_{int}^\gamma)/2 > 0$.
This corresponds to the preferred FeAFMMnFM ordering
Fe$_1^\uparrow$ Fe$_2^\downarrow$ Mn$_1^\uparrow$ Mn$_2^\uparrow$
and
explains the close competition with G-AFM ordering, in which both layers are AFM ordered
(see Fig.~\ref{AFMFMnanofig} and Table~\ref{magenergytab}).
The exchange coupling between $d^5$ Fe and $d^4$ Mn
takes place via superexchange through the bridging O.
For angles close to 180$^\circ$, strong $\sigma$ bonding favors FM ordering. 
However, as the Fe-O-Mn 
angles deviate from 180$^\circ$ through the oxygen octahedron rotational distortion,  
the admixture of $\pi$ bonding leads to an increasingly AFM character of the coupling~\cite{Goodenough55,Kanamori59,LeeAFM}.
This behavior can be seen in the dependence of the fitted values for $J_{int}$ 
on the Fe-O-Mn angle, plotted in Fig.~\ref{Jintanglefig}.

Within this nearest-neighbor Heisenberg model, 
we explored a wider range of possible magnetic orderings
for the R3c structure nanocheckerboard, in particular, 
orderings with lower translational symmetry than those
included in the first-principles investigation. 
The supercells considered included $2\times 2\times 2$ 
($p=8$ perovskite cells),
$2 \times 2 \times 4$
and $4 \times 2 \times 2$ 
($p=16$ perovskite unit cells). The Heisenberg model energies were computed for all 2$^p$ spin configurations in each supercell.

The lowest energy ordering found in this larger set of configurations is still 
the FeAFMMnFM ordering, with 
FM alignment of the Mn and AFM antialignment of the Fe along the Mn and Fe pillars, respectively,
and alternating FM and AFM $xy$ layers 
as in Fig.~\ref{AFMFMnanofig}.
The lowest-energy alternative magnetic state 
is a state in which one Mn per supercell in the FM $xy$ layer flips,
at an energy cost of $6.3$ meV/supercell.
The net magnetization for the resulting state decreases from 
$M_{GS} =3.8\mu_B$ per one Fe-Mn pair to $\frac{p-2}{p} M_{GS}$.

Within a mean field approximation with four effective fields,
two for the two Fe atoms and two for the two Mn atoms
in the unit cell of the BiFeO$_3$-BiMnO$_3$ nanocheckerboard,
the magnetic transition temperature
of the BiFeO$_3$-BiMnO$_3$ nanocheckerboard is $T_c = 406 K$.
This temperature is intermediate between the Neel temperature,
$T_N^{exp} = 643$~K,
of bulk BiFeO$_3$ and 
the Curie temperature, 
$T_c^{exp} = 105$~K,
of bulk BiMnO$_3$~\cite{Kiselev63,Chiba97,Sugawara65,Sugawara68,Faqir99}.

\section{Alternative Structures of the BiFeO$_3$-BiMnO$_3$ Nanocheckerboard}
\label{bfmoaltsec}

\begin{figure} [h]
\begin{center}
\includegraphics[scale=0.35,angle=-90]{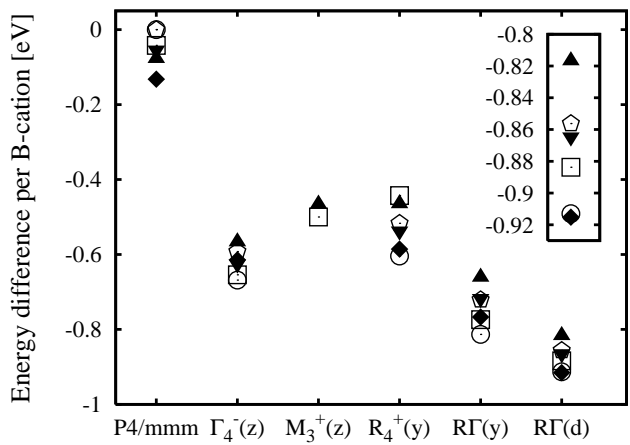}
\caption
{Structural energetics of BiFeO$_3$-BiMnO$_3$ nanocheckerboard.
Energy difference per perovskite cell (per B-cation)
for different magnetic orderings (see Fig.~\ref{AFMFMnanofig})
and for structural distortions (see Fig.~\ref{perovskitemodesfig} and Table~\ref{structurestab}).
Inset: zoomed view of the magnetic energies for the c-R3c structure.
FeAFMMnFM ordering (filled diamond)
competes with G-AFM ordering (open circle)
in the nanocheckerboard alternative structures.
}
\label{ediffdistbfmofig}
\end{center}
\end{figure}

\begin{table*}
\begin{center}
\caption{
 Low energy alternative and ground state (GS)
 structure of BiFeO$_3$-BiMnO$_3$ nanocheckerboard.
 Energy difference $\Delta E[eV/B-cation]$ 
 is calculated 
 for different magnetic orderings
 with respect to FeAFMMnAFM P4/mmm structure
 [as in Fig.~\ref{ediffdistbfmofig}].
 Insulating DOS band gap $\Delta$ (or metallic $m$ character),
 polarization $P$ estimated from the 
 nominal charges [Eq. (\ref{pointchargeP})],
 the in-plane $a$ and out-of-plane $c$ lattice constants
 [see perovskite cell in Fig.~\ref{checkerboardfig}],
 and oxygen-octahedron rotation angle $\Theta$
 are given for the lowest-energy magnetic ordering
 corresponding to each structural distortion.
} 
\label{BFMOtab}
{\small
\begin{tabular}{ccccc}
\\
\hline
\hline
Space group      & P4/mmm & c-P4mm & c-I4cm & c-R3c  \\
Modes	   & -      &$\Gamma_4^-(z)$ & $R_4^+$,$\Gamma_4^-(y)$ & $R_4^+$,$\Gamma_4^-([111])$ \\
\hline
 Mag. order     & FeAFMMnFM & G-AFM & G-AFM & FeAFMMnFM \\
$\Delta$  [eV]  & m      & 1.01   & 1.55   &  0.90 \\
 P $[\mu C/cm^2]$ & -      & 101.9  & 71.5   &  57.9 \\
 a/ c  [\AA]    &3.81/ 3.88 &3.66/ 4.60 &5.80/ 3.67  & 5.50/ 3.93  \\
 $\Theta [^{\circ}]$ &-       &-       &4.8     &7.2-20.3\\
\hline
$\Delta E$ (FeAFMMnFM)  & -0.132 & -0.615 & -0.767 & -0.915 \\
$\Delta E$ (G-AFM)      &  0.000 & -0.668 & -0.813 & -0.913 \\
$\Delta E$ (C-FIM)      & -0.042 & -0.654 & -0.774 & -0.884 \\
$\Delta E$ (FeFMMnAFM)  & -0.054 & -0.625 & -0.717 & -0.865 \\
$\Delta E$ (FeAFMMnAFM) &  0.000 & -0.592 & -0.721 & -0.856 \\
$\Delta E$ (FM)         & -0.077 & -0.567 & -0.661 & -0.817 \\
\hline
\hline
\end{tabular}
}
\end{center}
\end{table*}

The energies for 
various 
magnetic orderings and structural
distortions 
of the nanocheckerboard are shown in
in Fig.~\ref{ediffdistbfmofig}.
The structural parameters for each structure type are relaxed for each magnetic ordering.
The most energetically favorable
alternative structures, like the ground state FeAFMMnFM c-R3c structure, are polar 
and include oxygen octahedra rotations.

The polar distortion
in the alternative structures of the nanocheckerboard is quantified by 
the value of the polarization computed using nominal charges 
(Eq. (\ref{pointchargeP})),
that can be directly compared with nominal-charge polarizations in the structures of BiFeO$_3$
(cf. Tables~\ref{BFOtab} and~\ref{BFMOtab}).
As in BiFeO$_3$, there is a low-lying 
supertetragonal P4mm phase, with $c/a \sim 1.3$
and very large spontaneous polarization.
For the various structures considered, the polarization tends to
decrease as rotational distortion is introduced,
with the 
smallest value found in the c-R3c structure.

In the P4/mmm structure, the nanocheckerboard
is metallic,
while
a band gap opens with either polar or rotational distortion.
This behavior is similar to that of BiMnO$_3$ and BiFeO$_3$, 
which are metallic in the FM Pm$\bar{3}$m, or P4/mmm structures 
with a band gap opened and/or widened by distortion
(in FM BiMnO$_3$, only in the spin-down channel).

As can be seen in Fig.~\ref{ediffdistbfmofig},
the difference in energies between different structure
types is generally much larger
than the difference in magnetic energies for a given structure type.
The interesting feature of this figure is that the favored magnetic ordering is different for different structure types, switching between 
ferrimagnetic FeAFMMnFM  and 
antiferromagnetic G-AFM. This is in contrast to the case of bulk BiFeO$_3$ 
(see Fig.~\ref{ediffdistbfofig}), 
or bulk BiMnO$_3$ (see Fig.~\ref{ediffdistbmofig}),
in which the favored magnetic ordering does not change for different
structure types.

\section{Magnetic and Structural Transitions Driven by Anisotropic Strain}
\label{bfmostrainsec}

The sensitivity of the magnetic exchange couplings to the structure should produce changes in the magnetic ordering energies for perturbations that couple to the crystal structure, such as electric field, pressure and epitaxial strain. It is even possible that a structural perturbation could drive the system through a magnetic transition into an alternative low-energy ordering. Furthermore, the fact that in the nanocheckerboard the favored magnetic ordering is different for different structure types, discussed in the previous section, suggests that the magnetic ordering of the system could in principle be changed by a perturbation that changes the structure type, producing a novel magnetic-coupling response at the magnetic-structural phase 
boundary~\cite{Fennie06,Palova09short}. For example, 
it might be possible to drive the nanocheckerboard from its ferrimagnetic Fe\-AFM\-Mn\-FM
c-R3c GS with a nonzero magnetization
to a G-AFM c-I4cm state with zero magnetization.

We have explored this possibility for two types of epitaxial strain. 
First, 
we investigated the c-R3c phase with an isotropic epitaxial strain,
corresponding to an (110) matching plane.
Thus, the second and third lattice vectors of the 
$\sqrt{2}a \times \sqrt{2}a \times 2c$ supercell, 
along [-110] and along [001], are constrained to be
perpendicular with uniform scaling of the lattice constants 
$a=(1+s)a_0$ and $c=(1+s)c_0$,
where $a_0 = 5.50 \AA $ 
and $c_0=3.93 \AA$
are the unstrained lattice constants of FeAFMMnFM c-R3c GS.
In this case, there is no magnetic transition: 
the system remains FeAFMMnFM
from $s=0\%$ up to 
strain of $10\%$.

Second, we considered an anisotropic epitaxial strain, 
corresponding to a (001) matching plane, 
such that
the lattice constant along [110] is fixed to 
$\sqrt{2} \times a_0 = 5.52 \AA $,
while the lattice constant $a'$ along 
[-110] is elongated, with strain defined as $\frac{a'-a_0}{a_0}$.
$a_0= 3.9 \AA $ is chosen as it 
is the lattice constant of an
ideal perovskite cell with volume which is the average of that of bulk R3c BiFeO$_3$
($V_{BFO}= 59.28 \AA^3/B-cation$)
and bulk C2/c~\cite{Belik06,Belik07} BiMnO$_3$
($V_{BMO}= 59.41 \AA^3/B-cation$).

\begin{figure} [t!]
\begin{center}
\includegraphics[scale=0.55]{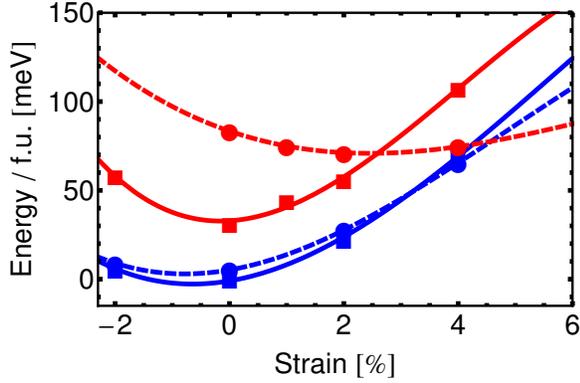}
\caption
{
Epitaxial-strain-driven magnetic transition in
BiFeO$_3$-BiMnO$_3$ nanocheckerboard.
Total energies of the FeAFMMnFM (solid)
and G-AFM (dashed line) magnetic orderings
in the 
c-R3c structure type (blue) 
and c-I4cm  structure type (red),
as a function of anisotropic in-plane tensile strain
(details in the text).
}
\label{magtransitionfig}
\end{center}
\end{figure}

The anisotropic epitaxial strain dependence of the energies of the c-R3c and c-I4m structures is presented in Fig.~\ref{magtransitionfig}.
At 0\% strain, the energy difference between FeAFMMnFM to G-AFM in the c-R3c structure is 
5 meV/B-cation; 
this differs slightly from the value for the relaxed structures reported 
in Table~\ref{BFMOtab}
due to the difference in lattice constants between the epitaxial constrained structure and the fully relaxed structure (the corresponding energy difference at 0\% strain for c-I4cm is greater because the difference in lattice constants is greater). At 3\% strain, there is a magnetic transition from ferrimagnetic FeAFMMnFM to this low-lying G-AFM phase, while the structure remains c-R3c. This arises from the modification of the exchange couplings by the structural changes produced by the changing epitaxial strain. 

With a further increase in strain, there is a transition from c-R3c to a c-I4cm phase at about 4.5\%. Since the favored magnetic ordering is G-AFM
in both structural phases, no magnetic transition accompanies the structural transition. However, this result does illustrate the feasibility of a epitaxial-strain-induced structural transition from one pattern of rotational distortions to another in this system; a coupled structural-magnetic phase boundary thus may be brought to light by future exploration of various epitaxial strain constraints.

\section{Role of the B-Site Cation Ordering in Magnetostructural Effect}
\label{bfmomagidealsec}

\begin{table*} 
\begin{center}
\caption{Calculated total magnetic energies and energy 
differences
in an ideal perovskite structure with lattice constant
$a_0=3.839\AA$
for various magnetic states in the checkerboard, 
rocksalt 
(oxygens are relaxed to accomodate their preferable positions),
and layered superlattice of BiFeO$_3$-BiMnO$_3$.
The checkerboard ordering shows a quasidegenerate spectrum
of magnetic energies,
whereas the rocksalt and layered superlattice
show larger gaps between the ground state (GS)
and the first alternative magnetic state.
Values of $U_{Fe}=U_{Mn}=5eV$ and $J_{Fe}=J_{Mn}=1eV$ are used in the first, 
and $U^{eff}_{Fe}=4eV$, $U^{eff}_{Mn}=5.2eV$ with $U^{eff}=U-J$ in the second column
(see Sec.~\ref{methodsec}).
} 
\label{tablemagenergiesU4}
{\small
\begin{tabular}{cccccc}
\\
\hline
\hline
 Checkerboard   & E [eV/B-cation]  & Rocksalt superlattice  & E [eV/B-cation] &  Layered superlattice   & E [eV/B-cation]\\
 magnetic state &  & magnetic state &   & magnetic state &  \\
\hline
FeAFMMnFM  & -35.04, -34.68 & FMFM      & -35.06, -34.66 &FeAFMMnFM & -35.11, -34.76 \\
  & $\Delta E$ [eV/B-cation] & & $\Delta E$ [eV/B-cation]& & $\Delta E$ [eV/B-cation] \\
FeAFMMnFM   & 0.000, 0.000 & FMFM       & 0.000, 0.000 &FeAFMMnFM   & 0.000, 0.000 \\
FM          & 0.022, 0.028 & FeAFMMnFM  & 0.044, 0.070 &FM          & 0.111, 0.097 \\
C-FIM       & 0.076, 0.113 & FeFMMnAFM  & 0.045, 0.055 & FeFMMnFM   & 0.136, 0.143 \\
FeAFMMnAFM  & 0.081, 0.084 & FM         & 0.065, 0.047 & FeAFMMnAFM & 0.135, 0.137 \\
G-AFM       & 0.114, 0.152 & AFMAFM     & 0.101, 0.084 & G-AFM      & 0.181, 0.219 \\
FeFMMnAFM   & 0.119, 0.129 & G-AFM      & 0.114, 0.079 & FeFMMnAFM  & 0.260, 0.257 \\
\hline 
\hline
\end{tabular}
}
\end{center}
\end{table*}

Quasidegenerate magnetic states
in the BiFeO$_3$-BiMnO$_3$ nanocheckerboard 
are a necessary ingredient for 
the observed magnetostructural effect
(cf. Fig.~\ref{ediffdistbfmofig} and Table~\ref{BFMOtab}),
where change in the magnetic ordering
is achieved by a perturbation
(e.g., epitaxial strain, as in Sec.~\ref{bfmostrainsec}]).
Here, we investigate the role of the cation-ordering geometry
in determining the spectrum of magnetic states; in particular,
this will show whether the quasidenerature spectrum
is unique to the checkerboard geometry.

Magnetic energies are computed in the ideal perovskite
structure of five systems:
the bulk parent BiFeO$_3$ and BiMnO$_3$,
the BiFeO$_3$-BiMnO$_3$ nanocheckerboard,
the BiFeO$_3$-BiMnO$_3$ layered (001) superlattice,
and the BiFeO$_3$-BiMnO$_3$ rocksalt structure
with Fe and Mn alternating in every other unit cell
((111) superlattice).
The results 
for the checkerboard,
the rocksalt, and the layered superlattice
are presented in Table~\ref{tablemagenergiesU4}.
Bulk BiFeO$_3$ and bulk BiMnO$_3$ exhibit
behavior similar to the (001) layered superlattice~\cite{Palova09short}.
In these three systems,
the difference in energy between the magnetic ground state
(G-AFM in bulk BiFeO$_3$, 
FM in bulk BiMnO$_3$,
and FeAFMMnFM in the (001) layered superlattice)
and the first alternative state is 
in the range $0.10-0.14 eV/B-cation$.
This spectral gap is sufficiently large that 
structural changes cannot lower the energy of an alternative state below that of the original magnetic ground state.

Indeed, for all structures considered
the lowest magnetic state
in bulk BiFeO$_3$ and bulk BiMnO$_3$
is G-AFM and FM, respectively
(see Figs.~\ref{ediffdistbfofig},~\ref{ediffdistbmofig}).
In the (001) superlattice,
we calculate magnetic energies for the G-AFM
and FeAFMMnFM magnetic states in two structural
distortions:
For l-I4cm (see Table~\ref{structurestab}),
we find 
$\Delta E = -0.504 eV/B-cation$ for G-AFM
and 
$\Delta E = -0.553 eV/B-cation$ for FeAFMMnFM
with respect to the FeAFMMnFM magnetic state in the ideal
perovskite cell (see Table~\ref{tablemagenergiesU4}).
For l-R3c,
we find
$\Delta E = -0.752 eV/B-cation$ for G-AFM
and 
$\Delta E = -0.761 eV/B-cation$ for FeAFMMnFM.
For both structural distortions considered,
the lowest energy magnetic ordering is FeAFMMnFM.

In contrast,
all magnetic states in the nanocheckerboard are quasidegenerate,
all are lower in energy than the lowest-energy states in the 
(001) superlattice and the bulk.
The rocksalt structure is an intermediate case: 
while the difference between the FMFM magnetic ground state
and the first alternative state
is $0.05 eV/B-cation$,
close to half of the spectral gap of the (001) superlattice
and bulk, all the states considered fall in the same low-energy 
window as for the checkerboard.
Therefore
it is much more likely that a structurally-driven transition
between the different magnetic states
could occur in the checkerboard,
or in the rocksalt structure,
than in the other geometries studied here.

The importance of the B-site cation geometry in the magnetic ordering energy spectrum can be qualitatively
understood from a simple Heisenberg model of the form given in Eq. (\ref{magneticE}), where we assume that the exchange couplings $J_{Fe}$, $J_{Mn}$ and $J_{int}$ are independent of cation geometry, thus being transferable from one geometry to the other.
We can approximately reproduce the magnetic ordering energies in the ideal perovskite structure of the checkerboard with  
the Mn-Mn interaction $J_{Mn}$ being strongly FM,
the Fe-Fe interaction $J_{Fe}$ being AFM and about half the strength,
and the Fe-Mn interaction $J_{int}$ being weakly FM.
Assuming the same values in the FeAFMMnFM GS
of the (001) layered superlattice, 
the high and medium-strength bonds are all satisfied ("happy" in the language of frustrated
magnetism) and the only unhappy bonds are weak bonds between the Mn and the 
opposite spin Fe in the adjacent layer (one bond per B cation). 
Thus this state is energetically clearly preferred over other orderings considered, 
which all involve a significant fraction of unhappy high and/or medium strength bonds, thus opening the observed gap in the magnetic energy spectrum. 
In contrast, in the checkerboard, the total fraction of high and medium-strength bonds is half that in the layered superlattice, and the alternative states are low in energy as they involve tradeoffs between a larger number of happy weak bonds and a smaller number of unhappy medium or high-strength bonds. 
Finally, in the rocksalt structure, all the nearest neighbor bonds are weak. This is consistent with the fact that all orderings considered are at low energies. However, a simple one-parameter model does not correctly account for the energetic order of the states in this range or the gap between the ground state ordering and the first alternative state, which would require a model including next-nearest neighbor interactions.

Indeed, the assumption of exact transferability used above is only semi-quantitatively valid. In particular, changes in B-site cation geometry result in 
relative energy shifts of
the Fe, Mn and O states and changes in the orbital wavefunctions, and thus
in changes to the wavefunction overlaps and energy denominators that
contribute to superexchange. This leads 
to different values of the magnetic couplings
$J_{Fe}$, $J_{Mn}$, or $J_{int}$ in the various geometries considered
(cf. Table~\ref{tablemagenergiesU4}).
In addition,
structural distortions modify these magnetic couplings,
as would be needed to explain the difference in the ordering of the
magnetic energies in Tables~\ref{BFMOtab} and~\ref{tablemagenergiesU4}.
However, the simple model does serve to give useful insight into this complex issue, 
and highlights the fact that the magnetic ordering spectrum is indeed very sensitive to the B-site cation arrangement.

\section{Discussion}
\label{discussionsec}

The experimental realization of the BiFeO$_3$-BiMnO$_3$ nanocheckerboard
would be challenging as
its formation energy is positive:
the combined total energies of R3c G-AFM ground state of bulk BiFeO$_3$
($E[BFO]=-35.079 eV/B-cation$)
and of the R3c FM  lowest energy structure of bulk BiMnO$_3$
($E[BMO]=-36.676 eV/B-cation$)
are lower than that of 
the c-R3c FeAFMMnFM ground state of the BiFeO$_3$-BiMnO$_3$ nanocheckerboard
($E[BFMO]=-71.694 eV/2 B-cations$).
Though
the BiFeO$_3$-BiMnO$_3$ nanocheckerboard is at best metastable,
there is indication from experiments
that fabrication of the BiFeO$_3$-BiMnO$_3$ nanocheckerboard with
square sizes on the order of a unit cell would not
be impossible with appropriate tuning of growth parameters. 
Growth of (001) BiMnO$_3$ on BiFeO$_3$ films has
recently been reported. In this study
post-annealing led to intermixing of the Fe and Mn,
with a concomitant increase in ferromagnetic $T_c$~\cite{Yang09}.
This experiment provides support for the first-principles observation that magnetic ordering in this system
is very sensitive to the B-site cation arrangement.
With regard to other film orientations,
(110) and (111) as well as (001) BiFeO$_3$ films have been 
successfully grown on oriented SrTiO$_3$ substrates~\cite{Li04,Chu07}.
For BiMnO$_3$, (111) and (001) oriented films have been grown with 
substrate vicinality~\cite{Sharan04,Son08}.
There should be no fundamental obstacle to analgous growth of (110)
oriented films of BiMnO$_3$. 
More generally, 
a combination of patterned substration,
possible masking,
layer-by-layer growth,
and carefully tuned growth parameters
could influence the deposition process enough to 
produce a checkerboard structure of BiFeO$_3$-BiMnO$_3$.

In order to make better contact with future experiment,
it is useful to consider magnetic ordering of larger-scale n$\times$n
BiFeO$_3$-BiMnO$_3$ checkerboards, 
where the lateral dimension of the BiFeO$_3$ and BiMnO$_3$ pillars
is n perovskite lattice constants.
Within each pillar, BiFeO$_3$ and BiMnO$_3$ regions should be G-AFM 
and FM respectively,
since this ordering is the most energetically favorable in 
the parent bulk structures. This is true even in the extreme case of n=1 discussed
in Sec.~\ref{bfmogsstrucsec}. 
The magnetic coupling constants 
$J_{Fe}$, $J_{Mn}$ and $J_{int}$ (Eq.~(\ref{Jeffvalues})),
obtained in Sec.~\ref{bfmomagr3csec} would allow
the construction of Heisenberg models 
to explore the magnetic ordering 
of these larger-scale checkerboards with the ideal perovskite structure,
where $J_{Fe}$ and $J_{Mn}$ connecting two atoms in the xy plane can
be taken to be equal to the coupling along the pillar, but we have not
pursued this farther here.
Generally speaking, we expect that 
the possibility of a structurally-driven magnetic transition should decrease as the 
lateral size of the BiFeO$_3$ and BiMnO$_3$ pillars
increases and the interface effects 
(Fe-Mn interactions)
become less important. 

On a technical note, the robustness of our calculated first-principles results 
has been checked by
using two different
implementations of LSDA+U with different parametrizations to compute
magnetic ordering energies in the ideal perovskite structure for the
checkerboard, the rocksalt cation ordering, and the layered superlattice
(Table VI).
The key results are the same for both implementations: 
the type of ground state magnetic ordering for each cation arrangement, the quasidegeneracy of the spectrum of magnetic energies
in the checkerboard,
the gap in the energy spectrum in the layered (001) superlattice,
and the intermediate character of the rocksalt ordering.

Finally we remark that our first principles calculations 
do not include spin-orbit coupling (SOC) that
is known to lead to 
weak ferromagnetism in BiFeO$_3$~\cite{Ederer05}.
Since the BiFeO$_3$-BiMnO$_3$ nanocheckerboard already has
a ferrimagnetic ground state without SOC,
inclusion of SOC may result in a slightly changed
value of
the total magnetization
and to small canting angles of the Fe and Mn spins;
these changes should not fundamentally affect the  
results presented here.
The addition of SOC to our present calculations is 
certainly worth pursuing in future work.

\section{Summary}
\label{summarysec}

In this paper, the structure and properties of 
an atomic-scale BiFeO$_3$-BiMnO$_3$ 
checkerboard were investigated	 	
using first-principles calculations and magnetic modeling.
This unusual heterostructure was found to have
properties distinct
from those of its bulk parent constituents, 
or those of (001) superlattices of these two
materials.  
We attribute this behavior to the magnetic frustration resulting from its
B-site cation geometry; 
this leads to a quasidegenerate manifold of magnetic 
states that can be switched through 
small applied external perturbations, resulting in an
unusual magnetostructural effect.  
The possibility of realizing this system in
the laboratory was discussed.  
This study of a two-component
nanocheckerboard should be considered as a proof-of-principle 
example, and we plan to study similar geometries 
on longer length scales to facilitate contact with future
experiments.

\section{Acknowledgments}
We thank 
E. Bousquet,
V. R. Cooper, 
M. Dawber, 
J. Driscoll,
C. Ederer,
C.-J. Eklund, 
C. J. Fennie, 
M. S. Hybertsen,
J. H. Lee, 
A. Malashevich, 
M. Marsman, 
J. B. Neaton, 
T. Nishimatsu, 
S. Patnaik,
O. Paz, 
D. R. Reichman,
N. A. Spaldin, 
D. Vanderbilt,
C. G. Van de Walle
and K. B. Whaley
for helpful discussions.
This work was supported in part by NSF MRSEC DMR-0820404, 
NSF NIRT-ECS-0608842 nad by the US Army Research Office 
through W911NF-07-0410.

\end{document}